\documentclass[10pt,final,journal,a4paper,twocolumn]{IEEEtran}

\usepackage{amsmath}
\usepackage{booktabs}
\usepackage{color}
\usepackage{epstopdf}
\usepackage{float}
\usepackage{graphicx}
\usepackage[utf8]{inputenc}

\usepackage{longtable}
\usepackage{multirow}
\usepackage{amssymb}
\usepackage{subcaption}
\usepackage{textcomp}
\usepackage{titlesec}
\usepackage{titling}
\usepackage{comment}
\usepackage{dblfloatfix}

\usepackage{url}

\usepackage[noabbrev]{cleveref}
\usepackage{tabulary}
\usepackage{caption}
\usepackage{gensymb}

\usepackage[numbers]{natbib}
\providecommand{\keywords}[1]
{{
  \small	
  \textbf{\textit{Keywords---}} #1
}}

\title{Computationally-Efficient Climate Predictions using Multi-Fidelity Surrogate Modelling}

\author{Ben Hudson, Frederik Nijweide, Isaac Sebenius\\
\small Computer Lab, University of Cambridge
\\\small\{bh511, fpjn2, iss31\}@cam.ac.uk}

\date{July 29th, 2021}

\begin{document}

\maketitle

\begin{abstract}
  Accurately modelling the Earth's climate has widespread applications ranging from forecasting local weather to understanding global climate change. Low-fidelity simulations of climate phenomena are readily available, but high-fidelity simulations are expensive to obtain. We therefore investigate the potential of Gaussian process-based multi-fidelity surrogate modelling as a way to produce high-fidelity climate predictions at low cost. Specifically, our model combines the predictions of a low-fidelity Global Climate Model (GCM) and those of a high-fidelity Regional Climate Model (RCM) to produce high-fidelity temperature predictions for a mountainous region on the coastline of Peru.
  We are able to produce high-fidelity temperature predictions at significantly lower computational cost compared to the high-fidelity model alone: our predictions have an average error of $15.62^\circ\text{C}^2$ yet our approach only evaluates the high-fidelity model on 6\% of the region of interest.
\end{abstract}

\hspace{10pt}
\keywords{Gaussian processes, Multi-fidelity modelling, Climate modelling, Earth observation}

\section{Introduction}

From predicting hourly weather forecasts to tracking global temperature changes over time, accurately modelling the Earth's climate is a pressing task with wide ranging impact. The climate science community has developed many models to simulate and predict weather and climatological dynamics. However, each model must balance geographical scale, spatial/temporal resolution, and computational cost. 
Global Climate Models (GCMs) model climate dynamics (e.g. temperature and wind) for the entire planet at once \cite{gfdl}. However, the computational cost of modelling the global climate is immense \cite{Armstrong}; thus, GCMs are restricted to a coarse spatial and temporal resolution. Regional Climate Models (RCMs) complement GCMs as they simulate the climate system over a particular region of the globe, but in much finer detail.
% The boundary conditions of these models are dictated by the output of the GCMs. 

Multi-fidelity surrogate models based on Gaussian processes (GPs) offer a unique opportunity to break the trade-off between simulation scale, resolution, and cost. These models can infer the high-fidelity predictions over a domain by learning the relationship between the low- and high-fidelity models based on a handful of samples from both models. Existing work suggests such techniques are promising for climate modelling \cite{Chang}.

In this paper, we evaluate the efficacy of multi-fidelity surrogate modelling to infer high-resolution temperature predictions in a mountainous, coastal region of Peru.
We analyse a dataset of pre-computed GCM (low-fidelity) and RCM (high-fidelity) predictions over this region.
Beginning with low-fidelity temperature predictions for the entire region,
we simulate running the high-fidelity model over square sub-regions of the region of interest, therefore acquiring the high-fidelity data on a batch-wise basis.
As there is a cost associated with obtaining this high-fidelity data (the computational cost of running the RCM), we aim to produce accurate high-fidelity temperature predictions while remaining within a certain budget.
We investigate if multi-fidelity models confer advantages over single-fidelity ones.
We also explore the impact of the choice of acquisition function on model performance, and if there are certain geographical regions over which it is especially important or unimportant to have high-fidelity data.

\section{Preliminaries}

\subsection{Gaussian processes}
Gaussian processes are a popular statistical tool for emulating black-box functions, such as complex simulations.
Intuitively, they can be understood of as a distribution over functions or an infinite collection of stochastic variables where any $N$ samples form an $N$-dimensional multivariate Gaussian distribution.

A Gaussian process is parameterised by a mean function $m(\mathbf{x})$, evaluated at each of $N$ input locations $\mathbf{x}$, and a kernel function $\kappa(\mathbf{x},\mathbf{x'})$ evaluated for each combination of input locations.
Many choices can be made for the kernel function, but the RBF kernel is often a popular choice \cite{martin_krasser_2020_4318528,Krasser2018}. The equation for this kernel is given by
\begin{equation}
 \kappa(\mathbf{x},\mathbf{x'})=\sigma^{2} \exp \left(-\frac{1}{2 l^{2}}\left(\mathbf{x}-\mathbf{x'}\right)^{T}\left(\mathbf{x}-\mathbf{x'}\right)\right),
 \label{eqn:rbf}
\end{equation}
where $\sigma$ scales the output of the kernel, and $\ell$ is the length scale, which determines how much the correlation decreases with the distance.

\subsection{Simulators and Emulators}
Simulators are useful for modelling climate and weather patterns (see \cite{STAN2014134}).
However, many (climate) simulations are very computationally intensive \cite{rasch2019overview}. Frequently, this makes them impractical for applications that require data to be updated often. However, it is possible to approximate the simulator's output with reasonable accuracy using statistical emulation, thus reducing the number of times the simulator must be evaluated. This is achieved by fitting a statistical model to the relation between inputs and outputs \cite{grow2014statistical}. Unlike simulators, which usually output single-valued functions over the input space, these models output probability distributions over the input space. For example, a Gaussian process-based emulator would output a normal distribution (characterised by a mean and standard deviation) for a point in the input space.

\subsection{Multi-fidelity Modelling}

When using simulators, there is often a trade-off between simulation cost and accuracy \cite{kennedy2000predicting}. ``Low-fidelity'' data can be produced easily using inexpensive and approximate simulation methods, yet often deviates significantly from reality. In contrast, ``high-fidelity'' data closely resembles the real-world system. This can be gathered from real-world measurements or computationally expensive simulations. 

Multi-fidelity modelling provides a useful framework for combining the accuracy of high-fidelity data with the low cost of low-fidelity data. In the simplest form, one can emulate high-fidelity data by scaling the low-fidelity data and adding an error term \cite{perdikaris2017nonlinear}. Mathematically,
\begin{equation}\label{eqn:mf}
 f_{\text {high }}(x)=f_{\text {err }}(x)+\rho f_{\text {low }}(x).
\end{equation}
When all the terms are independent Gaussian processes, we can perform mathematical operations, like addition, because the terms are multivariate normal distributions.

When a nonlinear relationship exists between high-fidelity and low-fidelity data, it can be modeled using a nonlinear information fusion  \cite{perdikaris2017nonlinear}, given by
\begin{equation}
 f_{\text {high }}(x)=\rho\left(f_{\text {low }}(x)\right)+\delta(x).
\end{equation}
These concepts behind multi-fidelity modelling can be extended to use more than two data sources, each having distinct costs and accuracies, including real-world data and other emulators \cite{damianou2013deep}.

\subsection{Acquisition Functions}

One fundamental problem in building a statistical emulator is deciding which new locations in the input space should be expensively evaluated. This is solved using an acquisition function: given a model with a known set of inputs $\mathbf{x}$ and an acquisition function $a(\mathbf{x})$, the next point to be evaluated is $\mathbf{x}_{n+1} = \mathrm{argmax}_{\mathbf{x} \in \mathbb{X}} a(\mathbf{x})$, where $\mathbb{X}$ denotes the total possible input space. Many acquisition functions exist, each with its own assumptions and optimization metrics \cite{forrester2008engineering,shahriari2015taking}. There are two acquisition functions that are relevant to our work.

\subsubsection{Model Variance}
This acquisition function selects sequential points based on the model's uncertainty, where each new selected point $x_{N+1}$ corresponds to the one with the highest variance, given by

\begin{equation}
    a_{MV,x} = \sigma^2(x).
    \label{eqn:model_var}
\end{equation}

\subsubsection{Integrated Variance Reduction}
Rather than choosing new points at which the model has the highest variance, the integrated variance reduction acquisition function aims to sample a new point which reduces the overall uncertainty of the model. More formally, one can approximate the integrated variance reduction as

\begin{equation}
    a_{IVR,x} = \frac{1}{|X|} \sum_{x_i \in X}\left[\sigma^{2}\left(x_{i}\right)-\sigma^{2}\left(x_{i} ; x\right)\right],
    \label{eqn:int_var}
\end{equation}

where $X$ is the set of test points used in the estimation in the calculation.

\section{Dataset}
\begin{figure}
        \centering
\includegraphics[width=0.8\linewidth]{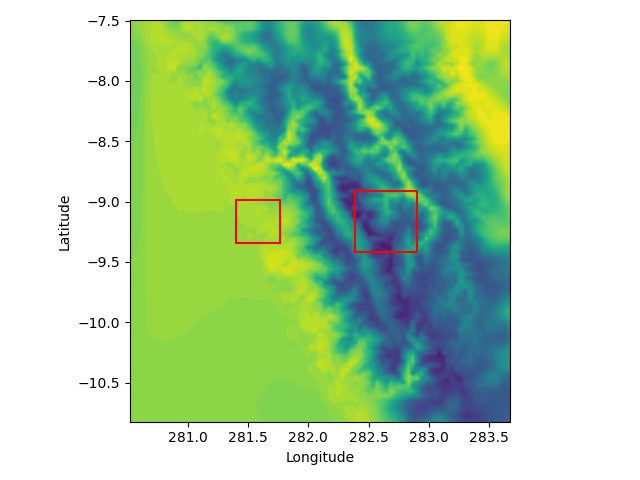}
\caption{The size of the sub-regions in which high-fidelity data is acquired. The smaller is about half the area of the larger region.}
\label{fig:grid_size}
     \end{figure}
     
We base our work on the multi-fidelity dataset provided by Hosking \cite{dataset}, which is comprised of low-fidelity and high-fidelity climate model outputs over a region of Peru (shown in \Cref{fig:roi} in \Cref{sec:appendix}). The high-fidelity data is available at $40\times$ higher spatial resolution than the low-fidelity data. Specifically, the following data is available, for each month from 1980 through 2018: 
\begin{itemize}
    \item \textit{High-fidelity temperature predictions}. The output from the RCM. This is the "target'' that we are interested in modelling.
    \item \textit{High-fidelity elevation data}. It is assumed that this elevation data remains constant over time.
    \item \textit{Low-fidelity temperature predictions}. The output from the GCM. This is inexpensive to compute, but predictions are at a coarser scale than its high-fidelity counterpart.
    \item \textit{Low-fidelity wind predictions}. The output from the GCM. It is available in North-South and East-West wind components.
\end{itemize}

\section{Methods}

We imagine having access to two climate models: a GCM, which can produce low-fidelity predictions of the temperature and wind speed quickly and inexpensively over the entire region of interest, and an RCM, which can produce high-fidelity predictions of the near-ground temperature using boundary conditions set by the GCM. However, the cost of running the RCM scales proportionally with the area covered by the simulation and is therefore prohibitively expensive to run over the entire region of interest. Instead, we run the RCM over several \textit{sub}-regions to remain within budget, and to combine these high-fidelity predictions with complete elevation data (and optionally low-fidelity GCM predictions) to infer the RCM's high-fidelity predictions over the remainder of the region of interest.

\Cref{fig:grid_size} shows two examples of sub-regions. Note that the sub-region on the right covers twice the area of the sub-region on the left. Thus, it would incur twice the cost to run the RCM over this region compared to the other.

\subsection{Models}
\begin{table}
    \centering\footnotesize
    \resizebox{\linewidth}{!}{
    \begin{tabulary}{0.8\linewidth}{lll}
    \toprule
        Model & Input & Output \\
        % \hline
        \midrule
        $HF$ & Latitude, Longitude, Altitude & HF Temp\\
        % \hline
        $LF\rightarrow{}HF$  & Latitude, Longitude, Altitude,  & HF Temp\\
        & LF Wind, LF Temp & \\
        % \hline
        $MF$ & Latitude, Longitude, Altitude & LF/HF Temp\\
    \bottomrule
    \end{tabulary}}
    \caption{Summary of the models evaluated.}
    \label{tab:models}
\end{table}
We compare three different models, summarised in \Cref{tab:models}. The first, $HF$ infers the high-fidelity temperature at a given location based only on the latitude, longitude and altitude of that location. The second $LF\rightarrow{}HF$ infers the high-fidelity temperature based on the latitude, longitude, altitude, low-fidelity temperature and wind speed at that location. Finally, $MF$ is a linear multi-fidelity model, as shown in \cref{eqn:mf}. This model infers the low-fidelity temperature and high-fidelity temperature at a given location based on the latitude, longitude, and altitude at that location.

\begin{figure*}[t]
    \centering
    \begin{subfigure}{0.49\linewidth}
        \centering
        \includegraphics[width=\linewidth]{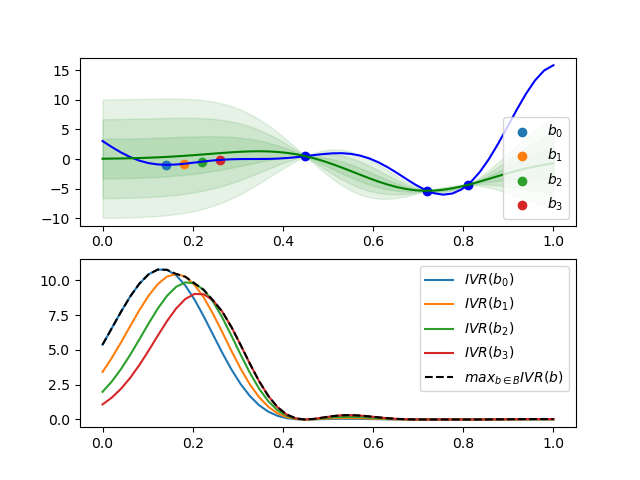}
        \caption{The variance reduction of integrating each point in the batch $B$, and our batch-wise acquisition function.}
        \label{fig:max_agg_strat_1}
    \end{subfigure}
    \hspace{0.1cm}
    \begin{subfigure}{0.49\linewidth}
        \centering
        \includegraphics[width=\linewidth]{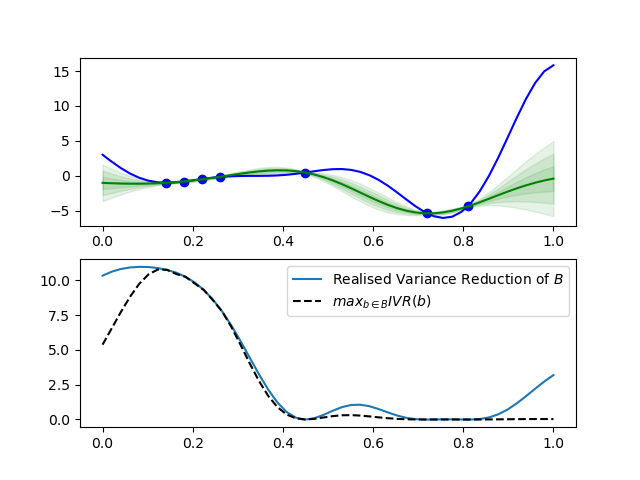}
        \caption{The realised variance reduction of integrating the batch $B$ compared to our batch-wise acquisition function.}
        \label{fig:max_agg_strat_2}
    \end{subfigure}
    \caption{Demonstration of our proposed batch-wise acquisition function, $a_{IVR,B,\max}$, on the Forrester function~\cite{forrester2008engineering} for a batch $B = \{b_0, b_1, b_2, b_3\}$. The test function is shown above and the integrated variance reduction is shown below.}
    \label{fig:max_agg_strat_demo}
\end{figure*}

\subsection{Batch Acquisition Function}

The crux of the problem is selecting where to run the costly RCM. This corresponds to selecting a batch of $n$ promising points to evaluate expensively, a task known as \textit{batch acquisition}. Batch acquisition is a well studied topic in statistical modelling, especially in Bayesian Optimisation~\cite{ginsbourger:hal-00260579,10.1145/3377930.3390154,pmlr-v124-jarvenpaa20a}. Some approaches jointly optimise these points' locations to maximise the acquisition function's value. In practice this is computationally intractable, so a heuristic is often used to select the points sequentially.

In our problem, the points in a batch are constrained to a small grid (see \Cref{fig:grid_size}). For each iteration, we would like to select the grid of points to evaluate expensively, such that the improvement of the model is maximised.

\subsubsection{Batch-wise Total Model Variance}\label{sec:mv_sum}

As a baseline method, we propose a batch-wise acquisition function extending the popular model variance function defined in in \cref{eqn:model_var}, based on Uncertainty Sampling. To do so, we evaluate the total model variance across a batch of points. This is given by

\begin{equation}
    a_{MV,B,\Sigma} = \sum_{b \in B}{\sigma^2(b)},
    \label{eqn:sum_agg_strat}
\end{equation}

where $B$ is a set of promising points.
When $B$ is a grid of points describing a sub-region, this heuristic selects the sub-region where the model variance is highest. As we are concerned with acquiring points in the high-fidelity model, we only evaluate the acquisition function over the high-fidelity component in the multi-fidelity case.

\subsubsection{Batch-wise Maximum Integrated Variance Reduction}\label{sec:ivr_max}

To improve on $a_{MV,B,\Sigma}$, we propose an extension of the integrated variance reduction acquisition function defined in \cref{eqn:int_var} for the batch acquisition scenario. This proposed heuristic evaluates the expected variance reduction when integrating each point in the batch at a set of test points and returns the maximum reduction across the batch for each test point. Mathematically, this is given by
\begin{equation}
    a_{IVR,B,\max} = \frac{1}{|X|}\sum_{x_i \in X}{\max_{b \in B}\left[\sigma^2(x_i) - \sigma^2(x_i;b)\right]}
    \label{eqn:max_agg_strat}
\end{equation}
where $B$ is a set of promising points and $X$ is the set of points at which to evaluate the variance.

This heuristic approximates the expected variance reduction of integrating an entire batch of points. 
\Cref{fig:max_agg_strat_demo} shows a demonstration of the heuristic in a simple scenario.
\Cref{fig:max_agg_strat_2} shows how the estimated variance reduction of integrating a particular batch $B = \{b_0, b_1, b_2, b_3\}$ compares to the realised variance reduction.  Note that $IVR(b) = \sigma^2(x) - \sigma^2(x;b)$. In the example presented, our heuristic underestimates the total variance reduction of integrating $B$ by 19\%. An analytic comparison of this heuristic to its exact counterpart is required to establish bounds on the estimation error. This is left to future work. When $B$ is a grid of points describing a sub-region this heuristic estimates the sub-region where the expected variance reduction of integrating the batch of points is maximised. Again, we only evaluate the acquisition function over the high-fidelity component in the multi-fidelity case.

\section{Experiments}
We evaluate the models in \Cref{tab:models}.
Each model acquires sub-regions of high-fidelity data according to the acquisition functions described in the previous sections (\Cref{sec:mv_sum} and \Cref{sec:ivr_max}).
We evaluate two differently-sized sub-regions, shown in \Cref{fig:grid_size}. The sub-regions are squares comprising of 121 points and 225 points respectively.
We limit the ``computational budget'' to a total of 500 points (excluding low-fidelity points) -- about 6\% of the region of interest. Therefore, approximately 4 small sub-regions or two large sub-regions can be acquired.

We initialise the $HF$ and $HF\rightarrow{}LF$ models with one randomly placed high-fidelity sub-region. We initialise the $MF$ model with 100 points from the low-fidelity model and one randomly placed high-fidelity sub-region.
We run each model 80 times and record the MSE of its predictions compared to the output of the RCM computed over the entire region of interest (the target).

We implemented the models using Emukit~\cite{paleyes2019emulation} and GPy~\cite{gpy2014}, and ran all experiments on virtual machines with 8 CPUs (Intel Cascade Lake generation) and 32 GB of RAM hosted on Google Cloud Compute Engine.

\section{Results \& Discussion}
\begin{figure*}
    \centering
    \begin{subfigure}[t]{0.24\linewidth}
        \centering
        \includegraphics[trim={2cm 0 2cm 0},clip,width=\linewidth]{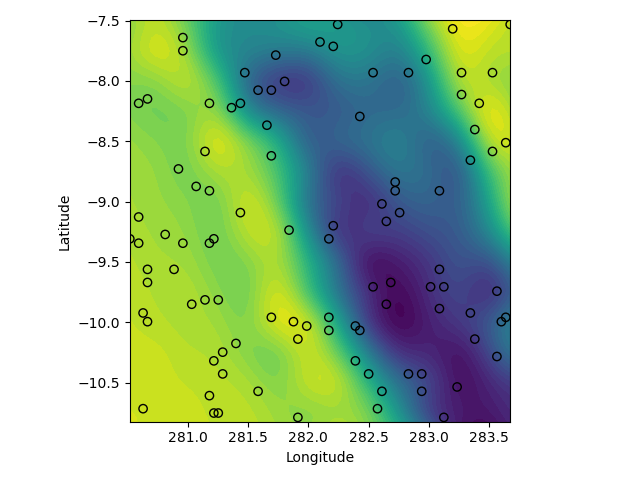}
        \caption{Low-fidelity\\training samples}
        \label{fig:lf_train_ex}
    \end{subfigure}
    \begin{subfigure}[t]{0.24\linewidth}
        \centering
        \includegraphics[trim={2cm 0 2cm 0},clip,width=\linewidth]{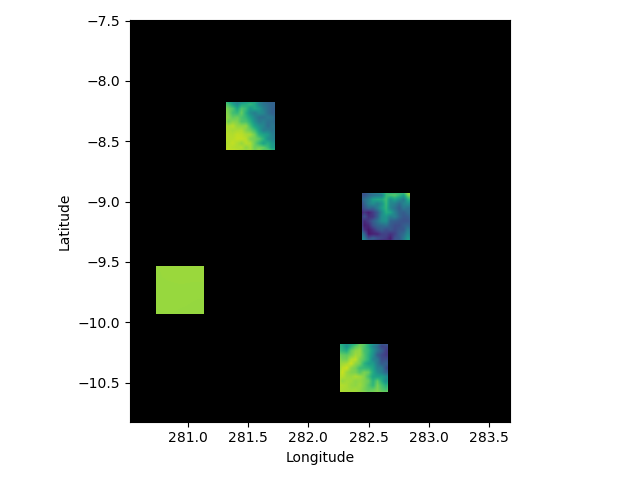}
        \caption{High-fidelity\\training samples}
        \label{fig:hf_train_ex}
    \end{subfigure}
    \begin{subfigure}[t]{0.24\linewidth}
        \centering
        \includegraphics[trim={2cm 0 2cm 0},clip,width=\linewidth]{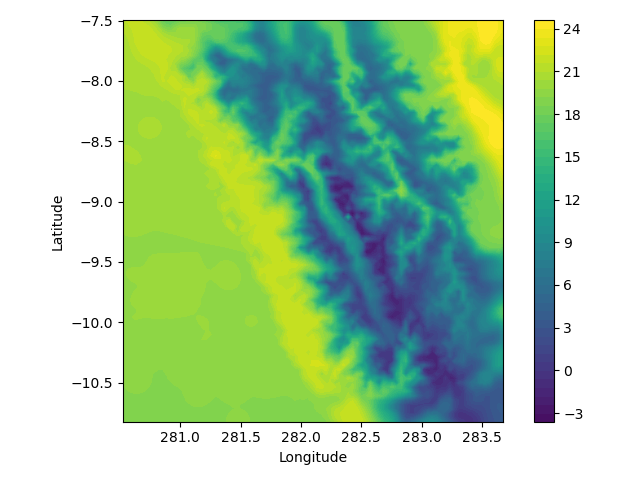}
        \caption{Inferred high-fidelity\\output}
        \label{fig:pred_ex}
    \end{subfigure}
    \begin{subfigure}[t]{0.24\linewidth}
        \centering
        \includegraphics[trim={2cm 0 2cm 0},clip,width=\linewidth]{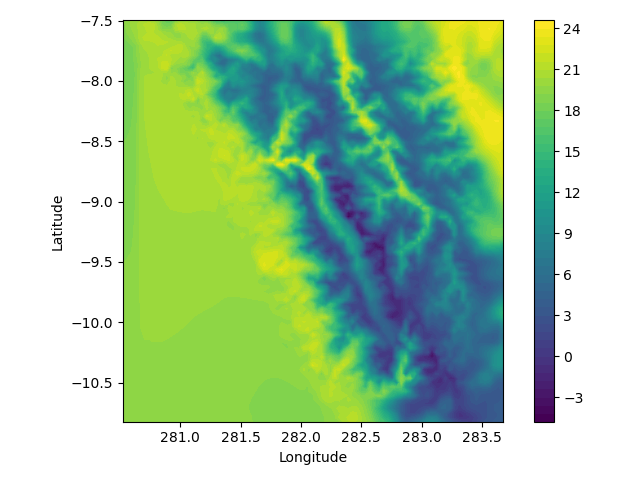}
        \caption{True high-fidelity output}
        \label{fig:true_ex}
    \end{subfigure}
    \caption{A sample of the multi-fidelity model using the proposed batch-wise acquisition function, $a_{IVR,B,\max}$, and acquiring small sub-regions of high-fidelity data, showing (a) the low-fidelity training samples from the GCM (b) the high-fidelity training sub-regions from the RCM (c) the inferred high-fidelity temperature predictions for the entire region of interest, and (d) the target high-fidelity temperature predictions from the RCM.}
    \label{fig:squares_ex}
\end{figure*}

\Cref{fig:squares_ex} shows an example of a prediction produced by the multi-fidelity model, including the low-fidelity points used, the high-fidelity sub-regions acquired, the statistical model's prediction, and the regional climate model's output over the entire region. The MSE for this example is $1.89^\circ\mathrm{C}^2$.

\Cref{tab:best_squares_sum} summarises the models' performance once the computational budget is reached.
The performance of all models as a function of the number of points acquired is shown in \Cref{fig:sweep_results} in \Cref{sec:appendix}.

\begin{table}
    \centering\footnotesize
    \resizebox{\linewidth}{!}{
    \begin{tabular}{lllrr}
\toprule
\multicolumn{3}{c}{Configuration} & \multicolumn{2}{c}{MSE} \\
            Model & Acquisition &  Region Size &  $\mu$ &  $\sigma$ \\
\midrule
          HF &     $a_{IVR,B,\max}$ & Small &   115.863 &   30.295 \\
          HF &     $a_{IVR,B,\max}$ & Large &   138.184 &   25.079 \\
          HF &    $a_{MV,B,\Sigma}$ & Small &   167.076 &   25.116 \\
          HF &    $a_{MV,B,\Sigma}$ & Large &   170.951 &   27.235 \\
LF$\rightarrow$HF &     $a_{IVR,B,\max}$ & Small &    94.302 &   53.988 \\
LF$\rightarrow$HF &     $a_{IVR,B,\max}$ & Large &   140.887 &   50.381 \\
LF$\rightarrow$HF &    $a_{MV,B,\Sigma}$ & Small &   139.977 &   38.321 \\
LF$\rightarrow$HF &    $a_{MV,B,\Sigma}$ & Large &   144.278 &   47.940 \\
               \textbf{MF} &     $\mathbf{a_{IVR,B,\max}}$ & \textbf{Small} &    \textbf{15.621} &   \textbf{18.109} \\
               MF &     $a_{IVR,B,\max}$ & Large &    52.450 &   58.641 \\
               MF &    $a_{MV,B,\Sigma}$ & Small &    20.145 &   21.276 \\
               MF &    $a_{MV,B,\Sigma}$ & Large &    46.713 &   47.814 \\
\bottomrule
\end{tabular}

    }
    \caption{Experimental results showing all combinations of model, acquisition function, and sub-region size. The multi-fidelity model using the batch-wise integrated variance reduction acquisition function and the small sub-region performs best.}
    \label{tab:best_squares_sum}
\end{table}

The multi-fidelity model using the batch-wise integrated variance reduction acquisition function, $a_{IVR,B,\max}$, and the small sub-region performs best, achieving an average MSE of $15.621^\circ\text{C}^2$.
The multi-fidelity model ($MF$) outperforms the single-fidelity models by a significant margin in all configurations. It could be that the small number of adjacent high-fidelity training points is insufficient to learn the correlation between the input and the high-fidelity output, but as the low-fidelity training points are reasonably uniformly distributed over the domain, they are sufficient to learn the correlation between the input and the low-fidelity output. Thus, the multi-fidelity approach could effectively bridge the learned relationships between input and low-fidelity output and between low- and high-fidelity output. 

Models using the smaller sub-region perform better than their counterparts. While both models acquire a similar number of points, the models using the small sub-region acquire more points according to the acquisition function (three small sub-regions, compared to only one large sub-region). As the acquisition function can direct where the points are acquired, it is unsurprising that the smaller sub-region models perform better.
However, the sub-region cannot be infinitely small as the RCM cannot be run over a single point. In a practical scenario, the minimum size of the sub-region should be dictated by a climate scientist or domain expert.

All models using the $a_{IVR,B,\max}$ acquisition function outperform their counterparts using the $a_{MV,B,\Sigma}$. The difference in performance is especially marked in the single-fidelity case, though improvements are observed in the multi-fidelity case as well.

\section{Conclusion}
We demonstrated that multi-fidelity surrogate modelling based on Gaussian processes can significantly reduce the cost of producing high-fidelity climate predictions. Our multi-fidelity model combines low-fidelity predictions from a GCM and a handful of high-fidelity sub-regions from an RCM to infer a high-quality prediction over an entire region of interest. Our model produces high-fidelity predictions with an average error of $15.62^\circ\mathrm{C}^2$ while only evaluating the RCM for 6\% of the region of interest. 
We demonstrated that surrogate modelling can be a useful tool for climate scientists, especially when used to emulate expensive simulators.

Additionally, we proposed a novel acquisition function, the $a_{IVR,B,\max}$, for the task of batch acquisition, demonstrating marked improvements over a batch model variance baseline. We used this to determine where to evaluate the RCM. 

\section{Future Work}

The work in this paper raises several interesting questions to pursue in future work.
There is a vast amount of low-fidelity data available, yet we initialised the multi-fidelity model using only about 1\% of it in order to train the model in a reasonable time. Is there an acquisition function that could be used to determine which \textit{low}-fidelity points to acquire to most improve the high-fidelity prediction? 

While we demonstrate that the $a_{IVR,B,\max}$ batch acquisition function improves performance over $a_{IVR,B,\Sigma}$, future work is necessary to analytically compare this heuristic to the exact integrated variance reduction of the entire batch.
Additionally, would an acquisition function that adapts the size of the sub-region it acquires make better use of the budget?
Finally, could our approach be extended to produce high-fidelity temperature predictions based on historical RCM predictions and both historical and current GCM predictions?
We believe that multi-fidelity surrogate modelling will play a key role in effectively modelling the Earth's changing climate as heterogenous observational data becomes more readily available.

\bibliography{bibliography}

\onecolumn

\appendix

\section{Supplementary Figures}\label{sec:appendix}

\begin{figure}[h]
     \centering
    \includegraphics[height=5cm]{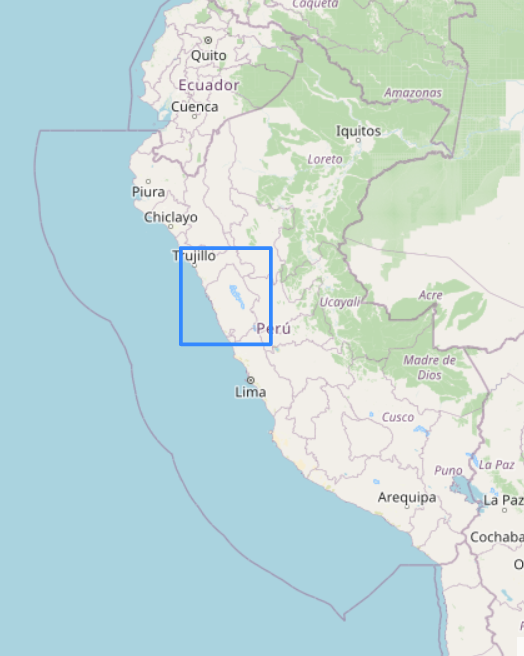}
    \caption{Region of interest}
    \label{fig:roi}
\end{figure}

\begin{figure}[h]
    \begin{subfigure}[t]{0.33\linewidth}
        \centering
        \includegraphics[width=\linewidth]{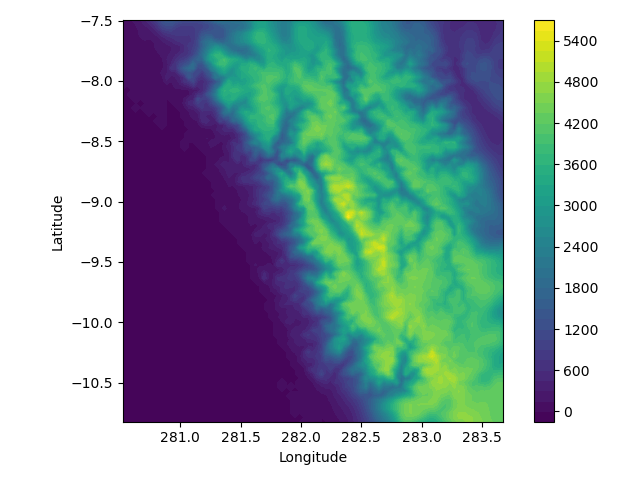}
        \caption{High-fidelity altitude}
        % \label{fig:lf_dataset}
    \end{subfigure}
    \begin{subfigure}[t]{0.33\linewidth}
        \centering
        \includegraphics[width=\linewidth]{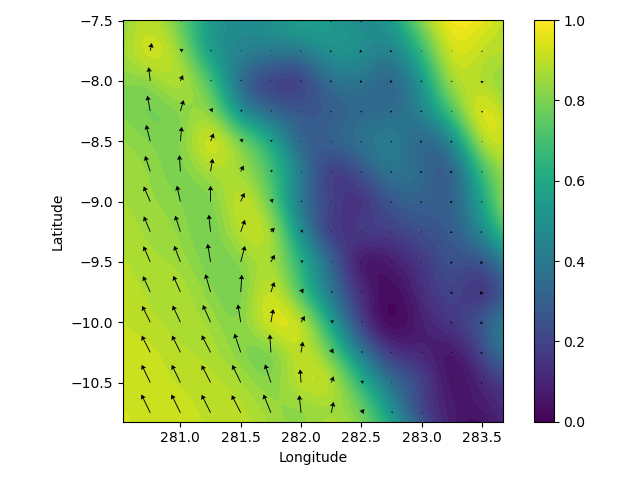}
        \caption{Low-fidelity temperature and wind}
        % \label{fig:lf_dataset}
    \end{subfigure}
    \begin{subfigure}[t]{0.33\linewidth}
        \centering
        \includegraphics[width=\linewidth]{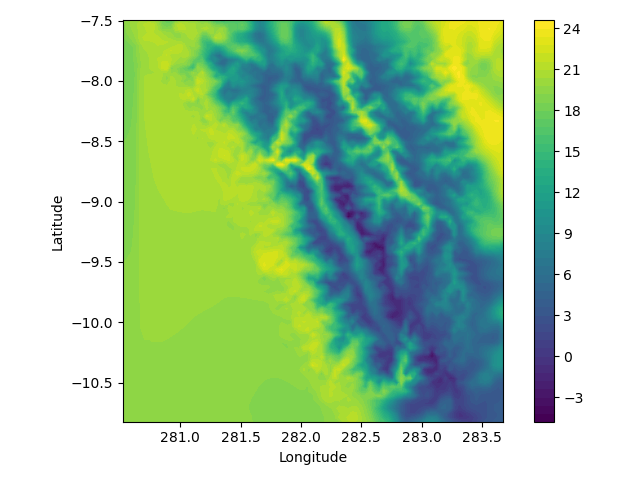}
        \caption{High-fidelity temperature}
        % \label{fig:lf_dataset}
    \end{subfigure}
    \caption{A visualization of the dataset studied.}
    \label{fig:dataset}
\end{figure}

\begin{figure}[h]
    \centering
    \begin{subfigure}{0.49\linewidth}
        \centering
        \includegraphics[width=\linewidth]{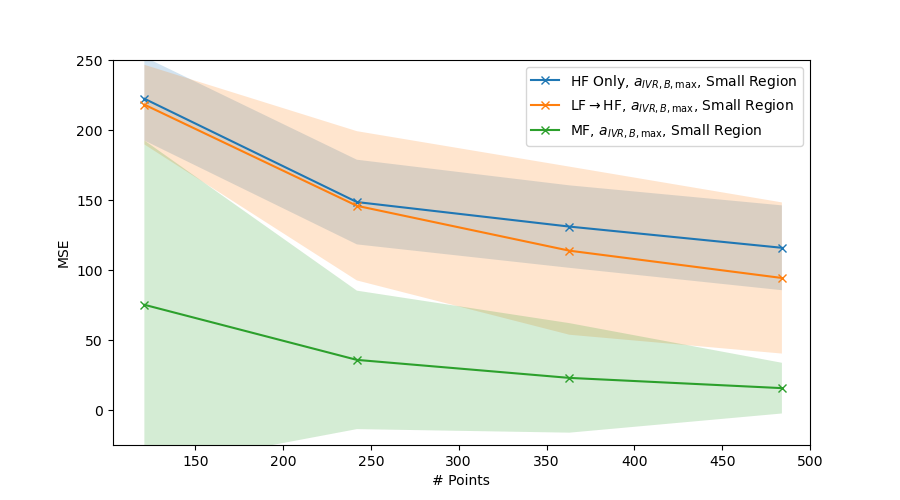}
        \caption{All models, $a_{IVR,B,\max}$, small region}
        \label{fig:ivr_sm}
    \end{subfigure}
    \begin{subfigure}{0.49\linewidth}
        \centering
        \includegraphics[width=\linewidth]{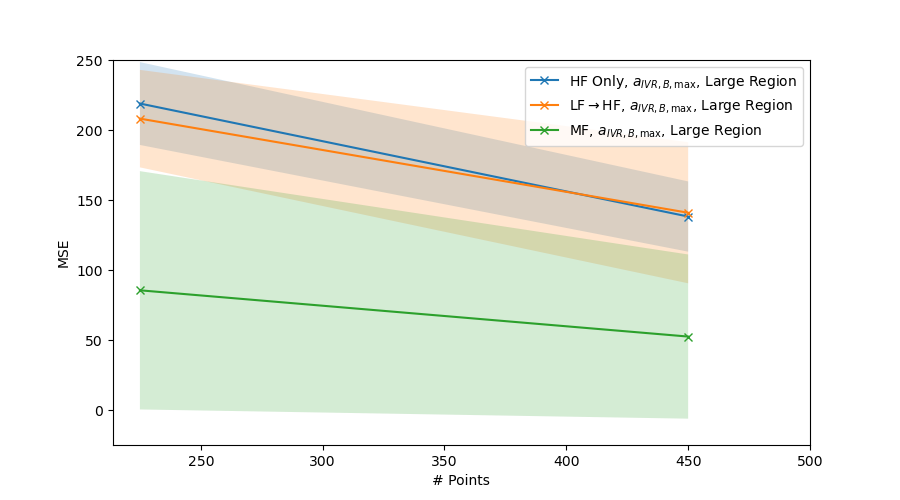}
        \caption{All models, $a_{IVR,B,\max}$, large region}
        \label{fig:ivr_lrg}
    \end{subfigure}
    \begin{subfigure}{0.49\linewidth}
        \centering
        \includegraphics[width=\linewidth]{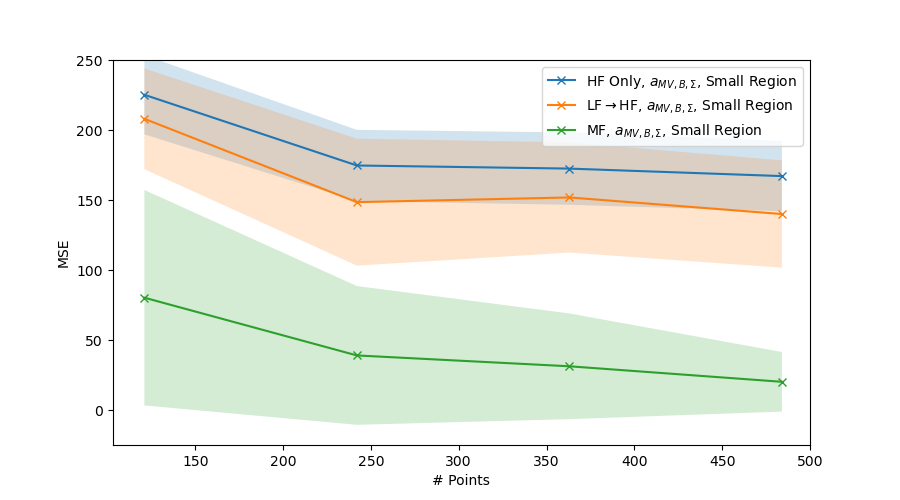}
        \caption{All models, $a_{MV,B,\Sigma}$, small region}
        \label{fig:mv_sm}
    \end{subfigure}
    \begin{subfigure}{0.49\linewidth}
        \centering
        \includegraphics[width=\linewidth]{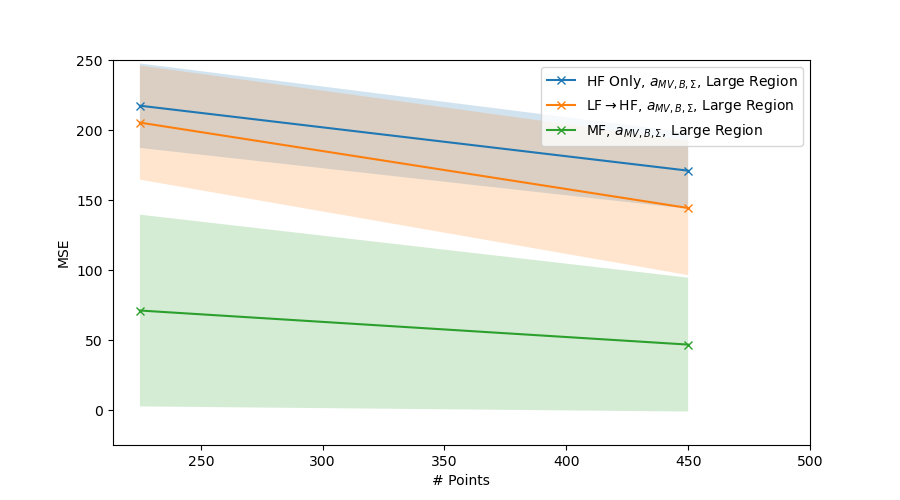}
        \caption{All models, $a_{MV,B,\Sigma}$, large region}
        \label{fig:mv_lrg}
    \end{subfigure}
        \begin{subfigure}{0.5\linewidth}
        \centering
        \includegraphics[width=\linewidth]{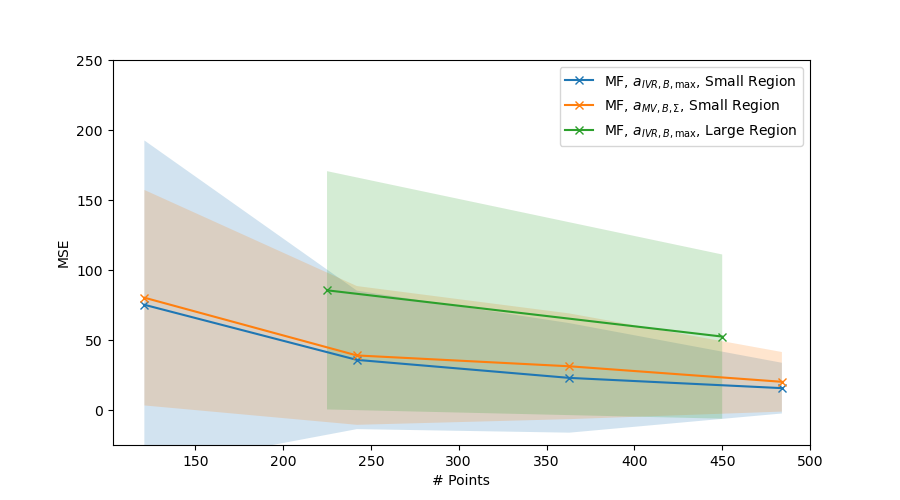}
        \caption{Best performing models}
        \label{fig:best_squares}
    \end{subfigure}
    \caption{Sweep results over model configurations}
    \label{fig:sweep_results}
\end{figure}

\end{document}